\documentclass[a4paper,12pt,numbers,sort&compress]{article}

\pdfoutput=1

\usepackage[paper=letterpaper,margin=1in]{geometry}

\usepackage{amsmath,amssymb,amsfonts,epsfig,cite,url}

\begin{document}
\newcommand{\todo}[1]{{ \bf ?????!!!! #1 ?????!!!!}\marginpar{$\Longleftarrow$}}
\newcommand{\nn}{\nonumber}
\newcommand{\tr}{\mathop{\rm Tr}}
\newcommand{\ch}{\rm Ch}
\newcommand{\comment}[1]{}

\newcommand{\ol}{\overline}
\def\av{\wedge^2 \cV}
\def\vv{\cV \otimes \cV^*}

\newcommand{\cN}{{\cal N}}
\newcommand{\cH}{{\cal H}}
\newcommand{\cO}{{\cal O}}
\newcommand{\cC}{{\cal C}}
\newcommand{\cD}{{\cal D}}
\newcommand{\cK}{{\cal K}}
\newcommand{\cT}{{\cal T}}
\newcommand{\cV}{{\cal V}}
\newcommand{\cE}{{\cal E}}
\newcommand{\cF}{{\cal F}}
\newcommand{\cR}{{\cal R}}
\newcommand{\IE}{\mathbb{E}}
\newcommand{\IF}{\mathbb{F}}
\newcommand{\IP}{\mathbb{P}}
\newcommand{\IQ}{\mathbb{Q}}
\newcommand{\IR}{\mathbb{R}}
\newcommand{\IC}{\mathbb{C}}
\newcommand{\IZ}{\mathbb{Z}}

\newcommand{\tmat}[1]{{\tiny \left(\begin{matrix} #1 \end{matrix}\right)}}
\newcommand{\mat}[1]{\left[\begin{matrix} #1 \end{matrix}\right]}
\newcommand{\diff}[2]{\frac{\partial #1}{\partial #2}}
\newcommand{\gen}[1]{\langle #1 \rangle}

\newtheorem{definition}{\bf DEFINITION}
\newtheorem{theorem}{\bf THEOREM}
\newtheorem{conjecture}{\bf CONJECTURE}
\newtheorem{proposition}{\bf PROPOSITION}

\def\theequation{\thesection.\arabic{equation}}
\newcommand{\setall}{\setcounter{equation}{0}
        \setcounter{theorem}{0}}
\newcommand{\setequation}{\setcounter{equation}{0}}

\begin{titlepage}

\begin{center}
{\Large \bf Calabi-Yau Spaces in the String Landscape}
\end{center}
\medskip

\vspace{.4cm}
\centerline{
{\large Yang-Hui He}
}
\vspace*{3.0ex}

\begin{center}
{\it
{\small
{
Merton College, University of Oxford, OX14JD, UK;\\
Department of Mathematics, City, University of London, EC1V 0HB, UK; and \\
School of Physics, NanKai University, Tianjin, 300071, P.R.~China\\
\qquad hey@maths.ox.ac.uk\\
}
}}
\end{center}

\renewcommand{\abstractname}{Article Summary}

\begin{abstract}
Calabi-Yau spaces, or K\"ahler spaces admitting zero Ricci curvature, have played a pivotal role in theoretical physics and pure mathematics for the last half-century. In physics, they constituted the first and natural solution to compactification of superstring theory to our 4-dimensional universe, primarily due to one of their equivalent definitions being the admittance of covariantly constant spinors.
Since the mid-1980s, physicists and mathematicians have joined forces in creating explicit examples of Calabi-Yau spaces, compiling  databases of formidable size, including the complete intersecion (CICY) dataset, the weighted hypersurfaces dataset, the elliptic-fibration dataset, the Kreuzer-Skarke toric hypersurface dataset, generalized CICYs etc., totaling at least on the order of $10^{10}$ manifolds.
These all contribute to the vast string landscape, the multitude of possible vacuum solutions to string compactification. 
More recently, this collaboration has been enriched by computer science and data science, the former, in bench-marking the complexity of the algorithms in computing geometric quantities and the latter, in applying techniques such as machine-learning in extracting unexpected information.
These endeavours, inspired by the physics of the string landscape, have rendered the investigation of Calabi-Yau spaces one of the most exciting and inter-disciplinary fields.
\end{abstract}

{\it Note: Invited contribution to the Oxford Research Encyclopedia of Physics, B.~Foster Ed., OUP, 2020}


\end{titlepage}

\tableofcontents

\section{Introduction and Motivation}
The coextensivity between mathematics and theoretical physics is very much a highlight of fundamental science in the XXth and XXIst centuries.
The two corner-stones of modern physics - {\it general relativity} (GR) describing the large structure of space-time, and {\it quantum field theory} (QFT) describing the elementary particles which constitute all matter - are well understood as mathematical in nature.
The former is the study of how the Ricci curvature of spacetime is induced by the energy-momentum tensor dictated by the Einstein-Hilbert action and the latter, of how particles of the Standard Model, furnishing representations of the Lie group $SU(3) \times SU(2) \times U(1)$, interact according to a gauge theory.
Both formalisms are tested by experiment to astounding accuracy, with the recent discovery of the Higgs boson and the detection of gravity waves being jewels in the crown of physics.

The brain-child of this tradition of mathematical unification, still dominating mathematical and theoretical physics as we enter the XXIst century, is string theory.
It is well-known by now that string theory is a ``theory of everything'' unifying GR and QFT in 10 space-time dimensions (or, equivalently, by dualities,  M-theory in 11 dimensions or F-theory in 12 dimensions). 
We must therefore account for $10-4=6$ ``missing'' dimensions.
Over the years, there has emerged a plethora of scenarios in the interplay between the physics of our four dimensions and the geometry of these 6 dimensions.
So great, however, is the number of such scenarios that it poses as one of the greatest theoretical challenges to modern physics.
This has come to be known as the ``vacuum degeneracy problem,'' where one is confronted with how to select {\it our} universe amidst an enormous number of solutions, what has come to be known as the {\bf string landscape}.

As a disclaimer from the outset, we will not cover the various issues of the {\it string landscape}, estimates of its vastness and the probabilities of finding our Standard Model therein. What will constitute the topic of the present article is the most standard and well studied class of solutions for the six extra dimensions: the so-called Calabi-Yau spaces. Our focus will be on the {\it Calabi-Yau Landscape}, a plethora of geometries in and of themselves.
In many ways, the aim is to guide a theoretical physicist through a lightning tour of some contemporary algebraic and differential geometry, with the jargons introduced in a pragmatic manner rather than through formalism.

We will take a chronological perspective, see how Calabi-Yau spaces arise in the physics, discuss how explicit constructions have been motivated by particle phenomenology and how a plethora of examples has been constructed under a long-term collaboration between mathematicians, physicists, and computer scientists.
First, however, let us  briefly set the scene  by introducing a classical mathematical problem dating back to Euler, Gauss and Riemann.
This digression  will help to set notation and definition, as well as to gain some intuition.

\paragraph{NOTE} This is a draft of a forthcoming article that has been accepted for publication by Oxford University Press in the {\it Oxford Research Encyclopedia of Physics, edited by Brian Foster}, due for publication in 2020.

%%%%%
\subsection{A Classical Problem in Mathematics}\label{s:classic}
\begin{figure}[t]
\[
\begin{array}{|c|c|c|}
\begin{array}{l}\includegraphics[width=1in,angle=0]{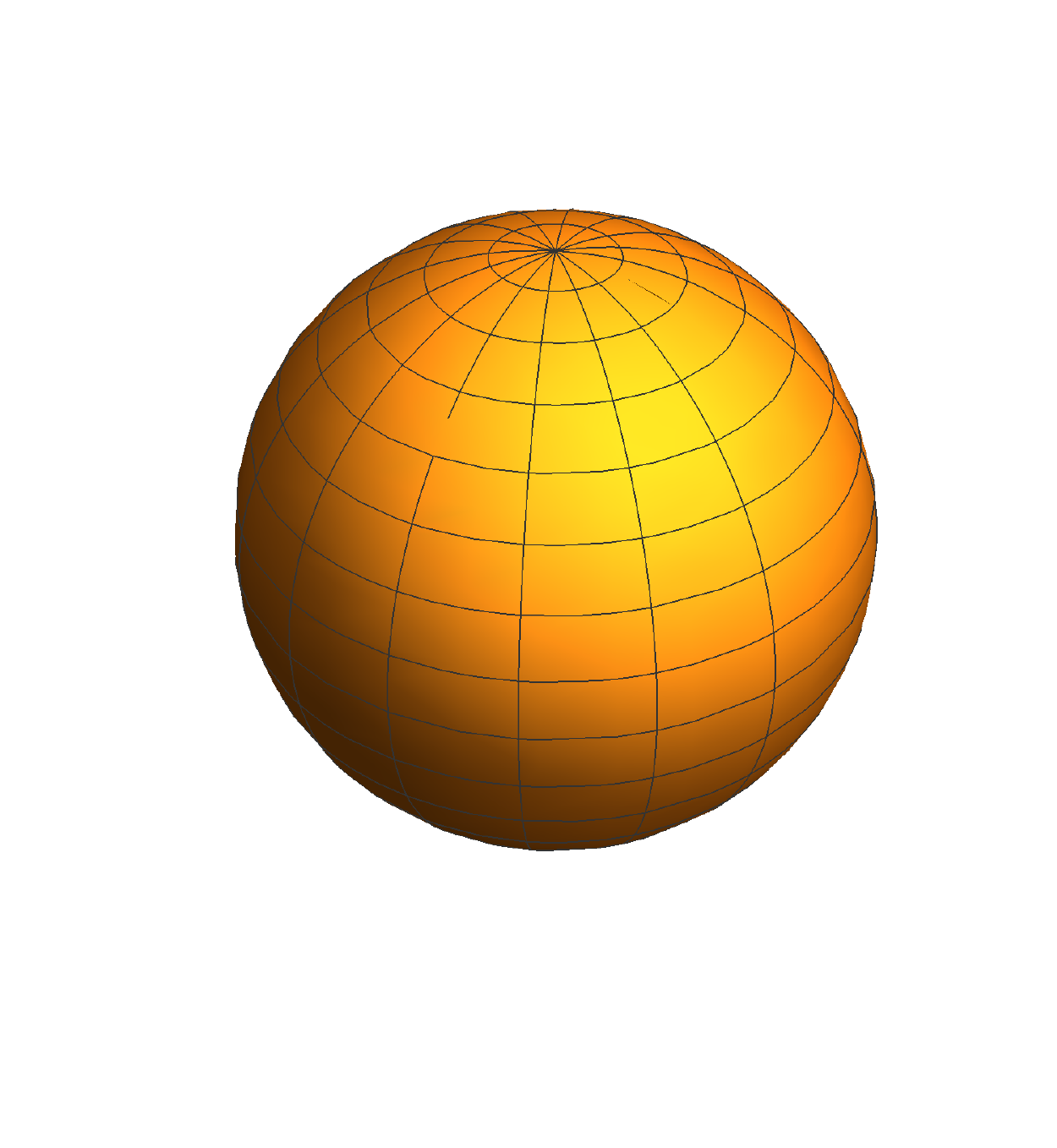}\end{array}
&
%ParametricPlot3D[{(3 + Sin[t]) Cos[f], (3 + Sin[t]) Sin[f], Cos[t]}, {t, 0, 2 Pi}, {f, 0, 2 Pi}, Boxed -> False, Axes -> False]
\begin{array}{l}\includegraphics[width=1in,angle=0]{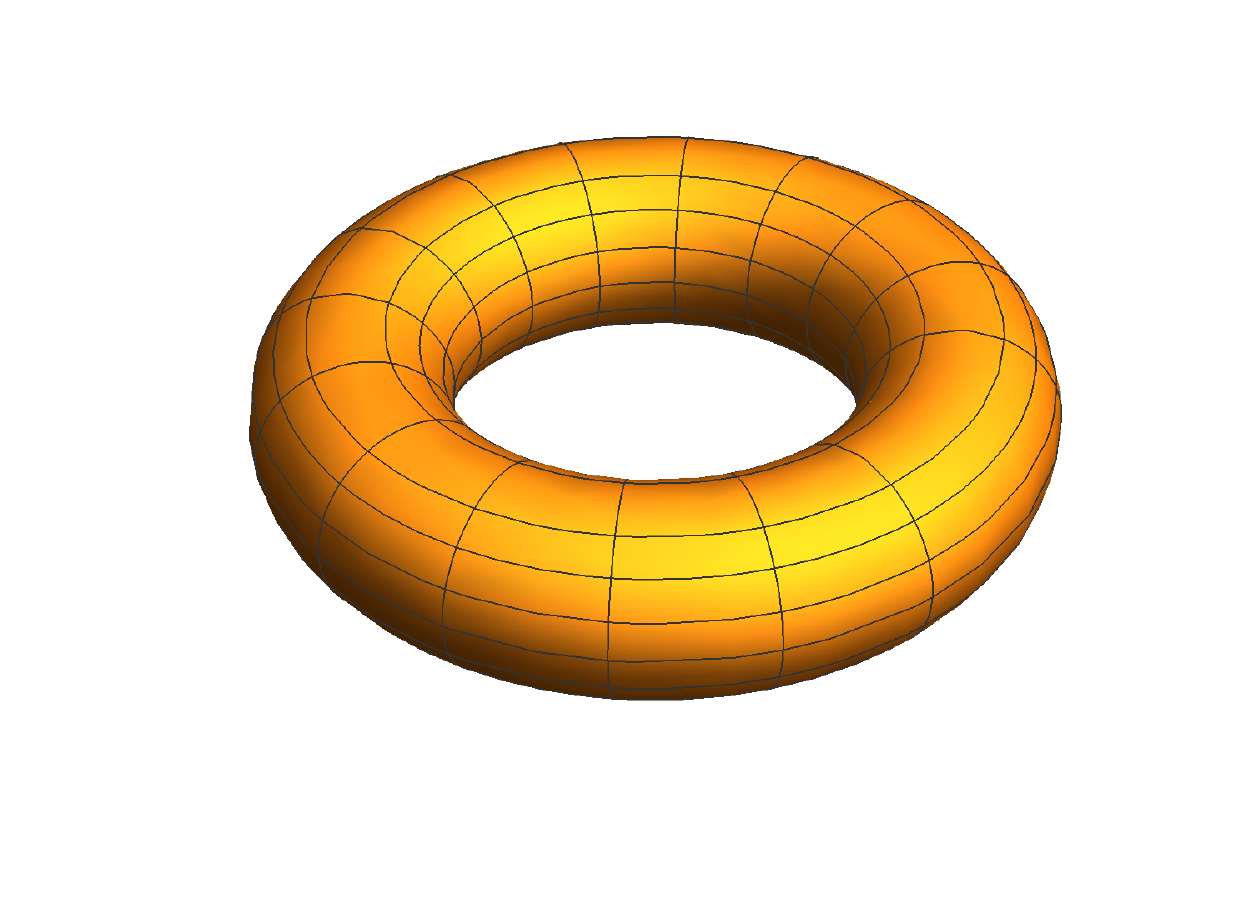}\end{array}
&
\begin{array}{l}\includegraphics[width=1.2in,angle=0]{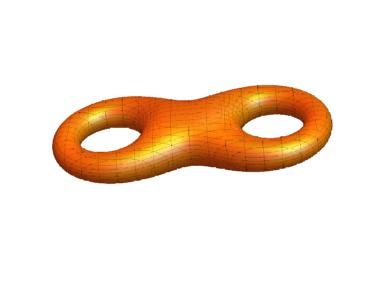}\end{array}
\begin{array}{l}\includegraphics[width=1.2in,angle=0]{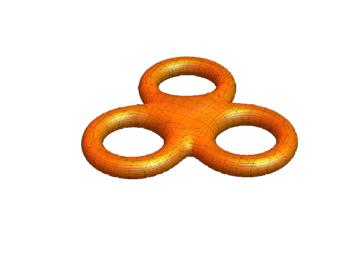}\end{array}
\begin{array}{l}\includegraphics[width=1.2in,angle=0]{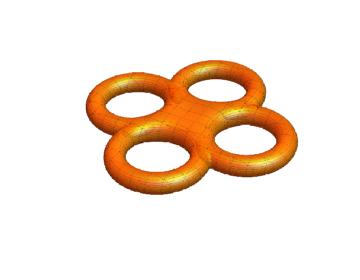}\end{array}
\ldots
\\ \hline
g(\Sigma) = 0 & g(\Sigma) = 1 & g(\Sigma) > 1\\ \hline
\chi(\Sigma) = 2 & \chi(\Sigma) = 0 & \chi(\Sigma) < 0\\ \hline
\mbox{Spherical} & \mbox{Ricci-Flat: CY$_1$} & \mbox{Hyperbolic} \\ \hline
+ \mbox{ curvature} & 0 \mbox{ curvature} & - \mbox{ curvature}
\end{array}
\]
\caption{{\sf
The trichotomy of (compact smooth) Riemann surfaces $\Sigma$ of genus $g(\Sigma)$, in terms of curvature, and related Euler number $\chi(\Sigma) = 2-2g(\Sigma)$.
Figures are generated on Mathematica \cite{mathematica}.
}
\label{f:RiemannSurface}}
\end{figure}

Consider a surface $\Sigma$ - we usually think of a sphere $S^2$ or the surface of a doughnut $T^2$ - and its possible {\bf topological types}, i.e., equivalences up to topology.
Restricting to the cases of smooth, compact (no punctures or boundaries) and orientable (nothing like Klein bottles or M\"obius strips) surfaces, the classic result is that the shapes which immediately came to mind, $S^2$, $T^2$, and those with increasing number of ``holes,'' are all there is:
any smooth, compact, orientable surface can be deformed continuously (topologically homeomorphic) to one of these.
We illustrate these surfaces at the top of Figure \ref{f:RiemannSurface}.
It is clear that a single non-negative integer classifies the topology of $\Sigma$, viz., the genus $g(\Sigma)$, referring to the number of ``holes.''
The quantity $\chi(\Sigma) = 2 - 2 g(\Sigma)$ is called the {\bf Euler characteristic} or Euler number.
We show both these quantities in the figure.

%%%%%%%%%%%%%%%%
%%%%%%%%%%%%%%
\comment{
building g > 1

torusImplicit[{x_, y_, z_}, R_, r_] = (x^2 + y^2 + z^2)^2 - 
   2 (R^2 + r^2) (x^2 + y^2) + 2 (R^2 - r^2) z^2 + (R^2 - r^2)^2;

build[n_] := 
  Module[{f, cp, polys, cartPolys, cartPolys1},(*implicit polynomial*)
   f = Product[
      torusImplicit[{x - 1.5 Cos[i 2 Pi/n], y - 1.5 Sin[i 2 Pi/n], z},
        1, 1/4], {i, 0, n - 1}] - 10;
   cp = ContourPlot3D[
     Evaluate[f == 0], {x, -3, 3}, {y, -3, 3}, {z, -1, 1}, 
     BoxRatios -> Automatic, PlotPoints -> 35, 
     MeshStyle -> Opacity[.2], 
     ContourStyle -> 
      Directive[Orange, Opacity[0.8], Specularity[White, 30]], 
     Boxed -> False, Axes -> False]];
}

As is often the case  in mathematics, it is expedient to work over the complex numbers $\IC$ rather than $\IR$ because of the algebraic closure of $\IC$ (every algebraic equation has solutions over $\IC$).
One can in fact think of $\Sigma$, instead of  as a 2-dimensional real object (a real surface), but as a one-complex-dimensional object (a complex curve).
For instance, $S^2$ is the complex plane $\IC$ with  the point at $\infty$ added.
Likewise $T^2$ is $\IC \simeq \IR^2$ quotiented by a lattice \footnote{This is seen as follows.
The plane $\IC$ can be tessellated by an infinite number of equal parallelograms $P$ the vertices of which constitute the lattice. Thus, the quotient first reduces the plane to a single $P$, the so-called fundamental domain.
Next, each pair of the opposite sides of $P$ is pasted to give a torus.}
What we have done is to endow $\Sigma$ with so-called {\bf complex structure}; every orientable surface $\Sigma$ can be thus equipped and is then called a Riemann surface or a (complex) algebraic curve \footnote{
It is rather confusing to the outsider that in algebraic geometry, because one usually works over $\IC$, a ``curve'' means a complex curve, of real dimension 2 and is really our familiar surface; likewise, a ``surface'' usually means a complex surface, of real dimension 4.
%Luckily, we will make very little mention of complex dimension 1 and 2 in this article since our main focus is on complex dimension 3.
}.
Hence, $\Sigma$ affords complex local coordinates $(z, \overline{z}) = (x + i y, \ x - i y)$ where $(x,y)$ can be thought of as the usual real coordinates.

In addition to complex structure, there is one more rich property one can put on $\Sigma$, the so-called {\bf K\"ahler structure}. If the metric 
 $g_{\mu \overline{\nu}}(z, \overline{z})$ comes from \footnote{Here, the indices $\mu, \overline{\nu}$ are just $z$ and $\overline{z}$ because we are in complex dimension 1; in complex dimension $n$, we will need $n$ pairs of coordinates $(z^\mu, \overline{z}^{\overline{\nu}}) $.} 
 a single scalar function, a (real) potential $\cK(z, \overline{z})$, via 
\begin{equation}\label{kahler}
g_{\mu \bar{\nu}}(z, \bar{z}) = \partial_\mu \overline{\partial}_{\overline{\nu}} K(z, \overline{z}) \ ,
\end{equation}
then, $\Sigma$ is not only complex, but is furthermore K\"ahler \footnote{The formal definition is the existence of a closed 2-form, the K\"ahler form $\omega$, associated with the Hermitian metric, which implies condition \eqref{kahler}.
}.
For Riemann surfaces, this is always possible.

With the introduction of the metric, it is natural to ask about curvature.
This is the classical theorem of Gauss-Bonnet: to relate curvature (differential gemetry) with the Euler number (topology).
In fact, we have a wonderful chain of equalities:
\begin{equation}\label{euler}
\chi(\Sigma) = 2-2g(\Sigma) =
\frac{1}{2\pi}\int_\Sigma R = 
 [c_1(\Sigma)] \cdot [\Sigma] = 
\sum\limits_{i = 0}^{\dim_{\IR} \Sigma = 2} (-1)^i b^i (\Sigma) \ .
\end{equation}
The first equality is the definition of the Euler number in terms of the genus.
The second relates it to the integration of the Gaussian curvature over $\Sigma$ and is the content of Gauss-Bonnet, relating the analytic concept of curvature to the topological concept of Euler number.
This is the epitome of classical {\it differential geometry}.

As we go further to the right, we enter {\it algebraic geometry}.
The third equality can be construed as a working definition 
\footnote{
Strictly, the Chern is defined for the tangent bundle $T_\Sigma$ of $\Sigma$; we will return to this point in \S\ref{genEmbed}.
Furthermore, we have written more formally as an inner product in intersection theory of the first Chern class $c_1$ (as an element, hence the square brackets, of the second cohomology $H^2(\Sigma)$) of the tangent bundle $T_\Sigma$ with (the second homology class $H_2(\Sigma)$) of $\Sigma$.
}
for the {\bf first Chern class} $c_1(\Sigma)$ of $\Sigma$: it is essentially the curvature $R$. 
Meanwhile, higher Chern classes $c_{i=2,3,\ldots}$ can be written as specific polynomials in $R$.
Importantly, these are topological invariants in that small perturbations of the metric do not change the value of the integral
$\int_\Sigma R$.

The fourth and last equality relates $\chi$ to the alternating sum \footnote{
Formally, $b^i$ is the ranks of the $i$-th homology group $H_i(\Sigma)$, which is also that of the $i$-th cohomology group $H_i(\Sigma)$.
} over the so-called {\bf Betti numbers $b^i$}.
This sum is the generalization of a familiar topological fact: for any convex polyhedron $P$ drawn on the sphere, with $V$ vertices (dimension 0), $E$ edges (real dimension 1) and $F$ faces (real dimension 2), we have that $V-E+F=2$ (take the cube, for example, we have $8-12+6=2$). 
For a convex polyhedron drawn on a genus $g$ Riemann surface, $V-E+F$ gives the Euler number $\chi$.
In general, the Betti numbers $b^i$ of a given space count independent dimension $i$ objects (formally, they are called algebraic cycles) therein.

Equation \eqref{euler} is a special case of the Atiyah-Singer Index Theorem, a deep result relating - as we proceed from left to right - topology to differential geometry to algebraic geometry to combinatorics.
From Figure \ref{f:RiemannSurface}, we see a natural trichotomy of Riemann surfaces: positive curvature (the sphere), zero curvature (the torus) and negative curvature ($g>1$). 
Let us keep in mind this boundary case of $T^2$: it is compact, smooth, Ricci-flat, K\"ahler and of complex dimension 1.

A central theme of modern geometry is to see how the various connections for Riemann surfaces/complex curves, on which we touched, generalize to beyond complex dimension 1/real dimension 2.
We now enter the realm of the geometry of {\bf manifolds}, i.e., shapes which are locally $\IR^n$, just like $\Sigma$ is locally $\IR^2 \simeq \IC$.
As expected, the situation is much more complicated.

Straight away, it is unfortunately {\it not} the case that every real $2n$-dimensional manifold can admit a complex structure and be turned into a complex $n$-fold \footnote{There are some further mild conditions, akin to the Cauchy-Riemann equations which are needed to define complex analytic functions.}
(note that throughout, we will adhere to the notation that an {\bf n-fold} is a complex manifold of complex dimension $n$ and real dimension $2n$).
Moreover, it is not true that every $n$-fold admits a K\"ahler structure.
Nevertheless, when a $2n$-manifold admits a complex structure and further, a K\"ahler structure, we have something akin to Figure \ref{f:RiemannSurface}, as we will now see.

%%%%%%%%%%%%%%%%%%%%%%%%
\subsection{A Modern Break-Through}
In 1954, E.~Calabi conjectured \cite{calabi} that for a (compact) K\"ahler manifold, the first Chern class uniquely governs 
\footnote{
More technically, the conjecture is as follows.
Let $(M,g,\omega)$ be a compact K\"ahler manifold and $R$ a $(1,1)$-form such that $[R] = [c_1(M)]$.
Then there exists a unique K\"ahler metric $\tilde{g}$ and K\"ahler form $\tilde{\omega}$ such that $[\omega] = [\tilde{\omega}] \in H^2(M; \IR)$ with
$R = Ric(\tilde{\omega})$, the Ricci form of $\tilde{\omega}$.  
}
the Ricci curvature of the K\"ahler metric \eqref{kahler}, much in the spirit of \eqref{euler}.
In particular, $c_1(M) = 0$ is the ``boundary'' situation where the metric is Ricci flat, much like our $T^2$ example for Riemann surfaces.

It took several decades for the existence part of this conjecture to be settled (the uniqueness is not too hard and follows from standard contradiction arguments).
S.-T.~Yau \cite{yau} proved the conjecture in 1978 using analysis of a PDE of Monge-Amp\`ere type.
The fact that Yau was immediately awarded the Fields Medal (1982) shows the importance of the result.
In honour of the two, the critical case of zero Ricci curvature K\"ahler manifolds (i.e., with $c_1 = 0$) are called {\bf Calabi-Yau manifolds}.
Henceforth, we denote a Calabi-Yau $n$-fold as CY$_n$.
Our familiar torus $T^2$ from the discussions in \S\ref{s:classic} is thus a (in fact, {\em the}) CY$_1$.
For the readers' convenience, we summarize the chain of specializations as follows:
\[
\mbox{Real $2n$-manifold}
\stackrel{\mbox{\tiny complex structure}}{\longrightarrow}
\mbox{Complex $n$-fold}
\stackrel{\mbox{\tiny potential $\cK$}}{\longrightarrow}
\mbox{K\"ahler $n$-fold}
\stackrel{\mbox{\tiny Ricci-flat}}{\longrightarrow}
\mbox{CYn}
\]

%%%
\subsection{Topological Quantities}
In \eqref{euler} we introduced the notion of the Betti number $b^i$ with $i$ ranging from 0 to the (real) dimension of the manifold.
As a complex number is paired with its conjugate, for a complex $n$-fold $X$, $b^i$ is refined into a doubled-indexed object $h^{j,k}$, counting the holomorphic (corresponding to the $z$-variables) and anti-holomorphic (corresponding to the $\overline{z}$-variables) pieces; these are so-called {\bf Hodge numbers}, satisfying
\begin{equation}
b^{ i = 0, \ldots, 2n } = \sum\limits_{j+k=i} h^{j,k}
\quad
\Rightarrow
\quad
\chi(X) = \sum\limits_{i = 0}^{2n} (-1)^i b^i = 
\sum\limits_{i = 0}^{2n} (-1)^i \sum\limits_{j+k=i} h^{j,k} \ .
\end{equation}
For smooth, compact $n$-folds, complex conjugation renders $h^{j,k} = h^{k,j}$ and something called Poincar\'e duality renders $h^{j,k} = h^{n-j, n -k}$.

Moreover, assuming that $X$ is connected (every point is linked to another by a path in $X$) renders $b^0 = h^{0,0} = 1$ and assuming that it is simply-connected (all closed paths can be retracted without breaking) forces $b^1 = h^{1,0} = h^{0,1}= 0$.
Finally, if $X$ is CY, then\footnote{
An equivalent definition for Calabi-Yau is that there is a unique holomorphic $n$-form, the volume form.
This unique form compels $h^{n,0}$ to equal 1.
}
 $h^{n,0} = h^{0,n} = 1$.

Putting all these constraints together, the point is that for (compact, smooth, simply connected) CY$_3$, there are only a few independent Hodge numbers, as illustrated by the following:
\begin{equation}\label{hodgediamond}
{\small
\begin{array}{ccccccc}
& & & h^{0,0} & & &  \\
& & h^{1,0} &  & h^{0,1} & & \\
& h^{2,0} & &  h^{1,1} & & h^{0,2} &  \\
h^{3,0} & & h^{2,1} & & h^{2,1} & &  h^{0,3} \\
& h^{2,0} & &  h^{1,1} & & h^{0,2} &   \\
& & h^{1,0} &  & h^{0,1} & &   \\
& & & h^{0,0} & & &   \\
\end{array}
=
\begin{array}{ccccccc}
& & & 1 & & &  \\
& & 0 &  & 0 & & \\
& 0 & &  h^{1,1} & & 0 &  \\
1 & & h^{1,2} & & h^{2,1} & &  1 \\
& 0 & &  h^{1,1} & & 0 &   \\
& & 0 &  & 0 & &   \\
& & & 1 & & &   \\
\end{array}
\leadsto
\begin{array}{c}
b^0\\
b^1\\
b^2\\
b^3\\
b^2\\
b^1\\
b^0\\
\end{array}
=
\begin{array}{c}
1\\
0\\
h^{1,1}\\
2h^{2,1} + 2\\
h^{1,1}\\
0\\
1\\
\end{array}
}
\end{equation}
Here, it is customary to write the Hodge numbers into a rhombic shape called the {\it Hodge diamond}.
We see that our CY$_3$ is characterized by only two integers $h^{1,1}$ and $h^{2,1}$, which count precisely the number of K\"ahler and complex structure deformations respectively.
We will address in \S\ref{s:finite} what other quantities completely characterize a Calabi-Yau manifold topologically.
Importantly, a consequence of \eqref{hodgediamond} and \eqref{euler} is that
\begin{equation}\label{eulerCY3}
\chi = \sum\limits_{i=0}^6 (-1)^i b^i = 2(h^{1,1} - h^{2,1}) \ ,
\end{equation}
which, interestingly, has physical significance, as we now address.

%%%%%%%%%%%%%%%%%%%%%%%%
\subsection{A Timely Relation to Physics}\label{s:physics}
What does all this beautiful mathematics have to do with physics?
First, Ricci-flat manifolds have long been known to physicists because they satisfy vacuum Einstein equations in GR.
Perhaps less familiar is the complex (and moreover, K\"ahler) condition.

This side of the story begins with Candelas-Horowitz-Strominger-Witten (CHSW) \cite{Candelas:1985en} in 1985, a culmination of the first string revolution
when Green-Schwarz \cite{Green:1984sg} cancelled anomalies and when the heterotic string of Gross-Harvey-Martinec-Rohm \cite{Gross:1984dd} naturally gave an $E_8$ gauge group to string theory.
Recall that $E_8$ is of particular significance, because of the following sequence of embeddings of Lie groups
\begin{equation}\label{embed}
SU(3) \times SU(2) \times U(1) \subset
SU(5) \subset
SO(10) \subset
E_6 \subset
E_7 \subset
E_8 \ .
\end{equation}
The first group $G_{SM} = SU(3) \times SU(2) \times U(1)$ is, of course, that of the Standard Model (SM).
One could also include the baryon-lepton symmetry $U(1)_{B-L}$ and write $G_{SM}' = SU(3) \times SU(2) \times U(1) \times U(1)_{B-L}$, in which case the above sequence of embeddings skips $SU(5)$.

The fact that $G_{SM}$ is not simple has troubled many physicists since the early days: it would be more pleasant to place the baryons and leptons in the same footing by allowing them to be in the same representation of a larger simple gauge group.
This is the motivation for the sequence in \eqref{embed}: starting from $SU(5)$, the gauge groups are simple and give archetypal {\bf grand unified theories} (GUTs), the most popular historically had been $SU(5)$, $SO(10)$ and $E_6$, long before string theory came onto the scene in theoretical physics.
Finally, throughout we will always consider {\bf supersymmetric} extensions of SM, such as the minimally supersymmetric standard model [MSSM] \footnote{
Whilst one might be skeptical of supersymmetry on grounds of there being still no experimental evidence, from a theoretical point of view it is definitely the right way of thinking.
An analogy with the real/complex number system is a fitting one.
Problems in mathematics are far better behaved over $\IC$ than over $\IR$ because the former is, as mentioned earlier, algebraically closed. Moreover, $\IC$ is in some sense (otherwise we will need to sacrifice commutativity and associativity) the unique extension of $\IR$ with this closure property.
Likewise, supersymmetry, by the theorem of Haag - {\L}opusza\'{n}ski - Sohnius \cite{Haag:1974qh}, is in some sense the unique extension for space-time symmetry under which QFTs are much better behaved. 
}.

In the compactification scenario, the 10-dimensional background is taken to be of the form $\IR^{1,3} \times X_6$ with $\IR^{1,3}$ our familiar space-time and $X_6$ some small curled-up 6-manifold too small to be currently observed directly, and possibly endowed with extra structure such as a vector bundle $\cV$ (we will return to this subject of bundles later).
CHSW gave the conditions for which the heterotic string, when compactified, would give a supersymmetric gauge theory in $\IR^{1,3}$ with a potentially realistic particle spectrum. 
Heuristically, one needs $X_6$ to be complex in order to have chiral fermions, and K\"ahler and Ricci-flat, to preserve supersymmetry and to solve vacuum Einstein's equations.
In other words, $X_6$ is a Calabi-Yau 3-fold \footnote{
It just so happened that Strominger, of CHSW, was sharing an office with Yau at the IAS, Princeton, as they arrived at their condition, not long after Yau got the Fields medal for CY. This  juxtaposition engendered the perfect discussion between the physicist and the mathematician.
This is a golden example of a magical aspect of string theory: it repeatedly infringes, almost always unexpectedly rather than forcibly, upon the most profound mathematics of paramount concern, and often proceeds to contribute to it.
}.

It should be emphasized that there are many possible solutions for $X_6$ as well as its associated further geometric structures.
Indeed, for the duality-equivalent 11-dimensional M-theory as well as 12-dimensional F-theory formulations of string theory, the compactification manifolds $X_7$ and $X_8$ allow even further possible solutions.
All these contribute to the vastness of the {\bf string landscape}, which we will not address here.
We will now see that even for CY$_3$, there is already a substantial multitude of possibilities.

%%%%%%%%%%%%%%%%%%%%%%%%%%%%%%%%%%%%%%%%%%%%%%%%%%%%%%%%%%
\section{Calabi-Yau Constructions and Databases}
As with any manifold, an important structure is its tangent bundle (for a sphere, think of this as the space of tangent planes).
For $X$, a CY$_3$, the tangent bundle $\cV = T_X$ has an extra property that it has $SU(3)$ structure, meaning that the group for parallel transport \footnote{
This is unlike for an arbitrary K\"ahler 3-fold, which has $U(3)$, or a Riemannian 6-manifold, which has $SO(6)$
} 
of vectors (the holonomy group), is $SU(3)$.
This $SU(n)$ holonomy is in fact another equivalent definition of a CYn.

Now, from Lie theory, we know that $SU(3)$ has commutant $E_6$ in $E_8$, i.e., a maximal commuting pair within the group $E_8$ is $SU(3) \times E_6$.
In relation to the compactification scenario in \S\ref{s:physics}, this means that from a CY$_3$ compactification we naturally have an (supersymmetric) $E_6$ GUT theory in $\IR^{1,3}$.
The particle content is readily determined from group theory.
More importantly, this in turn determines the vector bundle cohomology group that is associated with the particles.
The general paradigm is that 
\[
\mbox{``Geometry of $X_6$ $\longleftrightarrow$ physics of $\IR^{1,3}$.''}
\]
Thus is born the subject of {\bf string phenomenology}.

Specifically, we have that the decomposition of the adjoint 248 of $E_8$ breaks into $SU(3) \times E_6$ as $248\rightarrow (1,78)\oplus (3,27)\oplus (\overline{
3},\overline{27})\oplus (8,1)$. Thus the Standard Model particles, which in an $E_6$ GUT all reside in its $27$ representation, is associated with the fundamental $3$ of $SU(3)$. 
The 10-dimensional fermions are eigenfunctions of the Dirac operator, which splits into the 4-dimensional one and the 6-dimensional one on the CY$_3$.
The former gives the fundamental particles we see and the latter is governed by the geometry of $X$.
Later, in \S\ref{genEmbed}, we will see how this setup can be generalized to other GUT groups and to the SM.

Here, for example, it turns out that the 27 representation is precisely associated to $h^{2,1}(X)$ and the conjugate $\overline{27}$, to $h^{1,1}(X)$ (q.v.~\cite{GH}).
As promised at the end of \S\ref{s:physics}, the topological invariants of $X$ determine the particle content of the 4-dimensional physics.
%Similarly, the 1 representation of $E_6$ is associated with the 8 of $SU(3)$, and thus to $H^1(T_X \otimes T_X^*)$.
We summarize that for the fundamental fermions (the choice of which is anti-generation and which is generation is by convention):
\begin{equation}\label{gens}
\begin{array}{rl}
\mbox{generations of particles} & \sim h^{2,1}(X) \sim \mbox{Complex structure of $X$} \ , \\
\mbox{anti-generations of particles} & \sim h^{1,1}(X) \sim \mbox{K\"ahler structure of $X$} \ .
\end{array}
\end{equation}
An immediate {\it constraint} is that there should be 3 net generations \cite{PDG}, meaning that
\begin{equation}\label{TX3}
\left| h^{2,1}(X) - h^{1,1}(X) \right| = 3 \Rightarrow \chi(X) = \pm 6 \  ,
\end{equation}
where we have used \eqref{eulerCY3}.
Thus the endeavour of finding Calabi-Yau 3-folds with the property \eqref{TX3} began in 1986.
This geometrical ``love for threeness'' \footnote{Recently, independent of string theory or any unified theories, why there might be geometrical reasons for three generations to be built into the very geometry of the Standard Model has been explored \cite{He:2014oha}.
} has been more recently dubbed by Candelas et al.~as {\bf Triadophilia} \cite{Candelas:2007ac}.

Having extracted a mathematical problem from physical constraints, a practical quest in geometry resulted, and which has prompted some 40 years of research.
This was one of the first times theoretical physics gave a precise home-work problem to modern geometry:
{\it does there exist a (compact, smooth, simply connected) Calabi-Yau 3-fold with $\chi = \pm 6$}?
Indeed, how does one construct a CYn at all?

%%%%%%%%%%%%%%%%%%%%%%%
\subsection{The Quintic}
We are familiar with constructing shapes from Cartesian geometry.
For example, a quadratic equation in two real variables $(x,y)$ is a conic section, such as a circle, in $\IR^2$.
Thus, two variables with one polynomial constraint gives a $2-1=1$ dimensional real manifold in an ambient $\IR^2$.
This simple idea is the beginnings of {\bf algebraic geometry}.

Since we are looking for complex manifolds, we simply construct them as the zero-locus of multivariate polynomials in complex variables.
We recall from the introduction that a Calabi-Yau 1-fold is nothing but a Riemann surface of zero curvature, viz.~the torus $T^2 = S^1 \times S^1$.
This can be realized as a cubic in two complex variables given by  the so-called Weierstra\ss\ equation: 
$
T^2 \simeq \{x,y \in \IC | y^2 = x^3 - g_2 x - g_4\} \subset \IC^2
$,
where $g_{2,4}$ are complex constants \footnote{One can check by writing out $(x,y)$ in their real and imaginary parts, and the Weierstra\ss\ equation becomes 2 real constraints. We can then solve this numerical and plot the result to see a torus emerge.}.  

Now, we need compactness, which means we will need to include the point at infinity, where $(x,y) = (\infty, \infty)$.
One can do this by so-called {\it projectivization} where instead of $\IC^2$,  we introduce one more complex coordinate, $z$ such that the point $(x,y,z) \in \IC^3$ is to be identified with $\lambda (x,y,z)$ for any non-zero $\lambda \in \IC$.
This scale-invariance brings the point at infinity to a finite point, rendering the resulting ambient $\IC^2$ and the subsequent torus compact.
What we have done is to construct, from $\IC^3$ with coordinates $(x,y,z)$, the complex projective space $\IC\IP^2$ with scale-invariant (or so-called {\bf homogeneous}) coordinates $[x:y:z]$.
Formally, we can define $\IC\IP^n$ from $\IC^{n+1}$ with coordinates $(z_0, z_1, \ldots, z_n)$ via the quotient by the equivalence relation $\sim$ as
\begin{equation}\label{cpn}
 \IC\IP^n := \IC^{n+1} \backslash \{ \vec{0} \} \bigg/  (z_0,z_1,\ldots,z_n) \sim  \lambda  (z_0,\ldots,z_n)  \ , \qquad
  \lambda \in \IC \backslash \{0\} \ .
\end{equation}
Note that we take out the origin $\{ \vec{0} \}$ because it is a fixed point under $\sim$ and would give rise to a singularity; $\IC\IP^n$ is a smooth complex manifold of complex dimension $n$ with $n+1$ homogeneous coordinates.

In summary, we can construct a CY$_1$ as the zero-locus of a homogeneous cubic polynomial (the Weierstra\ss\ equation)
inside $\IC\IP^2$
\begin{equation}\label{weierstrass}
\{[x:y:z] | y^2z = x^3 - 4g_2 xz^2 - g_4z^3\} \subset \IC\IP^2 \ ,
\end{equation} 
where the homogeneous coordinates $[x:y:z]$ means that the point $(x,y,z)$ is to be identified with $\lambda(x,y,z)$ for any non-zero $\lambda$.

In a nutshell, {\em algebraic geometry is the study of how complex manifolds arise as zero-locus of multivariate complex polynomials in projective spaces}. When there is only a single polynomial, it is called a {\bf hypersurface}.
In general, there could be many polynomials defining the manifold as an {\it algebraic variety}.
For hypersurfaces, the dimension is simply that of the ambient space minus one. For instance, in the above example, the $T^2$ is of complex dimension $2-1=1$, as required.
Luckily, loci of homogeneous polynomials in complex projective space are guaranteed to be K\"ahler manifolds, so we are already half-way there.

This construction, of having a degree $n+1$ homogeneous polynomial in $\IC\IP^n$ is indeed valid in general: one can show that if the number of projective coordinates, here $n+1$, equals the degree of the hypersurface, this implies the vanishing of the first Chern class.
In other words, the hypersurface defined by a homogeneous polynomial of degree $n+1$ in $\IC\IP^n$ is a Calabi-Yau $(n-1)$-fold.
Thus we arrive at our first, and perhaps most famous, example of a Calabi-Yau 3-fold: the quintic hypersurface in $\IC\IP^4$.
There are many degree 5 monomials (the number of these different monomials is roughly the complex structure) one could compose of 5 coordinates, the most well-studied is the so-called Fermat quintic:
\[
Q := \{ x_0^5 + x_1^5 + x_2^5 + x_3^5 + x_4^5 = 0 \} \subset
\IC\IP^4_{[x_0:x_1:x_2:x_3:x_4]} \ .
\]
What are the topological numbers of $Q$? It turns out \footnote{
In this case, one can heuristically obtain the 1 as being descended from the K\"ahler class of the ambient $\IC\IP^4$ and the 101 as follows: $h^{2,1}$ counts the inequivalent complex deformations,
there are ${5+5-1 \choose 5} = 125$ degree 5 monomials in 5 variables, subtract this by the action of linear redefinition of variables which amounts to $(5^2 - 1) = 24$, and then by an overall scaling gives us $125 -24 - 1 = 101$.
In general, however, this naive counting does not work.
} that
$h^{2,1}(Q) = 101$ and $h^{1,1}(Q) = 1$ so that $\chi(Q) = 2(1-101) = -200$ and this is quite far from $\pm 6$.

%%%%%%%%%%%%%%%%%%%%%%%%%%%
\subsection{The CICY Database}
To continue to address the question raised in equation \eqref{TX3}, an algorithmic generalization of the construction for the quintic was undertaken: instead of a single $\IC\IP^n$, what about embedding a collection of (homogeneous) polynomials into a product $A$ of projective spaces?
For further simplification,  consider only {\it complete intersections} which means the optimal case where the number of equations is 3 less than the dimension of the ambient space $A$, so that each polynomial reduces exactly one degree of freedom.

In other words, let $A = \IC\IP^{n_1} \times \ldots \times \IC\IP^{n_m}$, of dimension $n = n_1 +n_2 + \ldots + n_m$ and each having homogeneous coordinates $[x_1^{(r)}:x_2^{(r)}:\ldots:x_{n_r}^{(r)}]$ with the superscript $(r) = n_1, n_2, \ldots, n_m$ indexing the projective space factors.
Our CY$_3$ is then defined as  the intersection of $K = n-3$ homogeneous polynomials in the coordinates $x_j^{(r)}$.
Clearly this is a generalization of the quintic, for which $r=m$, $n_r = 4$ and $K=1$ and more generally a hypersurface is thus trivially a complete intersection with 1 defining polynomial.

The Calabi-Yau condition of the vanishing of $c_1(T_X)$ generalizes analogously to the condition that for each $r=1, \ldots, m$, we have $\sum\limits_{j=1}^K q^{r}_{j} = n_r + 1$.
Succinctly, this information can be written into an $m \times K$ configuration matrix (to which we could adjoin the first column, designating the ambient product of projective spaces, for clarity; this is redundant because one can extract $n_r$ from one less than the row sum):
\begin{equation}\label{cicy}
X = 
\left[\begin{array}{c|cccc}
  \IC\IP^{n_1} & q_{1}^{1} & q_{2}^{1} & \ldots & q_{K}^{1} \\
  \IC\IP^{n_2} & q_{1}^{2} & q_{2}^{2} & \ldots & q_{K}^{2} \\
  \vdots & \vdots & \vdots & \ddots & \vdots \\
  \IC\IP^{n_m} & q_{1}^{m} & q_{2}^{m} & \ldots & q_{K}^{m} \\
  \end{array}\right]_{m \times K \ ,}
\quad
\begin{array}{l}
K = \sum\limits_{r=1}^m n_r-3 \ , \\
\sum\limits_{j=1}^K q^{r}_{j} = n_r + 1 \ , \ \forall \; r=1, \ldots, m \ .
\end{array}
\end{equation}

For example, $[5]$, or $[4|5]$, denotes the quintic.
Two more immediate examples are
\begin{equation}\label{egCICY}
S = \mat{1&1\\3&0\\0&3} \ , \qquad
\tilde{S} = \mat{1&3&0\\ 1&0&3\\} \ .
\end{equation}
The first is called the Schoen Manifold 
%while in the physics community, the second has come to be known as the Tian-Yau manifold.
and the second, the Yau-Tian.
Specifically, the configuration $S$ means that the ambient space is $\IC\IP^1 \times \IC\IP^2 \times \IC\IP^2$, of dimension 5.
We check that since there are 2 polynomials (columns), $5-2=3$ gives a CY$_3$.
The first column $(1,3,0)$ means that the first polynomial is linear in $\IC\IP^1$ and cubic in the first $\IC\IP^2$ while having no dependence on the second $\IC\IP^2$. The second column is likewise defined.  
Another lesson, as can be induced from $S$ and $\tilde{S}$, is the rather cute fact that the transpose gives a new valid configuration, though the pair can be of wildly different geometry.
We often attach the topological numbers as $X^{h^{1,1}, h^{2,1}}_\chi$ for completeness, so we can write, for instance, $Q^{1,101}_{-200}$, $S^{19,19}_0$ and $\tilde{S}^{14,23}_{-18}$ .
Importantly, the Chern classes and the Euler number can be read off the matrix configuration explicitly.
\comment{
We have that $c_1(T_X) = 0$ and moreover,
\begin{equation}\label{chernCICY}
c_2^{rs}(T_X) = \frac12 \left[ -\delta^{rs}(n_r + 1) + 
  \sum_{j=1}^K q^r_j q^s_j \right] \ , \quad
c_3^{rst}(T_X) = \frac13 \left[\delta^{rst}(n_r + 1) - 
  \sum_{j=1}^K q^r_j q^s_j q^t_j \right] \ , 
\end{equation}
where we have written the coefficients of the total Chern class
$c = c_1^r J_r + c_2^{rs} J_r J_s + c_3^{rst} J_r J_s J_t$ explicitly, with $J_r$ being the K\"ahler form in $\IP^{n_r}$.
The triple-intersection form $d_{rst} = \int_X J_r \wedge J_s \wedge J_t$ is a totally symmetric tensor on $X$ and the Euler number is simply $\chi(X) = d_{rst} c_3^{rst}$.
}
The individual terms $(h^{1,1},h^{2,1})$ in \eqref{eulerCY3}, however, cannot be deduced from the configuration matrix directly\footnote{
This is one of the short-comings of the index theorem: the integral of curvature and the intersection of the Chern classes give only the alternating sum (Euler number) in (co-)homology, but not the individual terms.
}.

Manifolds defined by \eqref{cicy} were considered and explicitly constructed by Candelas et al.~\cite{cicy} (q.v.~H\"ubsch's classic book \cite{hubschbook}) in the early 1990s and were affectionately called CICYs (complete intersection Calabi-Yau manifolds).
Classifying these above matrices, up to geometrical equivalence, was thus a relevant affair for physics: one could for instance read off the Euler number to see whether any of them had magnitude of 6.
Obviously it was also relevant for mathematics: until then, creating datasets of algebraic varieties was not the style of questions for pure mathematicians. 
The combinatorial problem for these integer matrices turned out to be rather non-trivial and one of the most powerful super-computers then available was recruited.
This was perhaps the first time when heavy machine computation was done for the sake of algebraic geometry.

In all, CICYs were shown to be finite in number, a total of 7890 inequivalent configurations, with a maximum of 12 rows, a maximum of 15 columns, and all having entries $q_j^r \in [0,5]$.
There are 266 distinct Hodge pairs $(h^{1,1},h^{2,1}) = (1,65), \ldots, (19,19)$,  giving 70 distinct Euler numbers $\chi \in [-200,0]$.

Unfortunately, none of the 7890 had $\chi = \pm 6$. While this was initially disappointing, it was soon realized that circumventing this problem gave rise to the resolution of another important physical question.
A freely acting order-3 symmetry was found on $\tilde{S}$; the freely acting is important, because it means that the quotient $S' = \tilde{S} / \IZ_3$ is also a smooth CY$_3$, albeit not a CICY.
For such smooth quotients, the Euler divides (the individual Hodge numbers do not and it turns out that we have $(h^{1,1},h^{2,1}) = (6,9)$ for $S'$) and $S'$ became the first three-generation manifold.

Now, the quotienting is crucial for another reason.
In the sequence \eqref{embed}, we have focused on GUTs. What about the Standard Model itself? It so happens that one standard way of obtaining $G_{SM}= SU(3)\times SU(2) \times U(1)$ from any of the GUT groups is precisely by quotienting.
Group theoretically, this amounts to finding a discrete group whose generators can be embedded into the GUT group, so that the commutant is the the desired $G_{SM}$.

Geometrically, this is the action of the {\bf Wilson Line}, where a CY$_3$ with non-trivial fundamental group admits a non-trivial loop which, coupled with the discrete group action, decomposes the $E_8$ to $G_{SM}$.
In our example above, $\tilde{S}$ has trivial fundamental group $\pi_1$ but the quotient $S'$ has, by construction, $\pi_1(S') \simeq \IZ_3$, thereby admitting a $\IZ_3$ Wilson line. This, for the early models, can be used to break the $E_6$ GUT down to the Standard Model.
Not surprisingly, the manifold $S'$ became central to string phenomenology in the early days \cite{Greene:1986ar}.

%%%%
\subsubsection{Recent Developments}\label{genEmbed}
As mentioned, the compactification scenario can in general involve Calabi-Yau manifolds as well as extra structures thereon, such as fluxes and bundles.
Indeed, CHSW gave a more general set of solutions in terms of the CY$_3$ $X$ as well as a field strength $F$.
Specifically, the set of conditions, known as the {\it Hull-Strominger System} \cite{Hull,Strominger:1986uh}, for the 
low-energy low-dimensional theory on $\IR^{1,3}$ to be a supersymmetric gauge theory are (rather schematically as we will not delve into the details of equations):
\begin{enumerate}
\item $X_6$ is complex;
\item The (Hermitian) metric $g$ (as well as its Ricci curvature $R$) on $X_6$ and field strength $F$ satisfy 
\begin{enumerate}
\item $\partial \overline{\partial} g = i \tr F \wedge F - i \tr R \wedge R$;
\item $(\partial + \overline{\partial}) ( g \  \Omega \wedge \overline{\Omega} ) = 0$, where $\Omega$ is a holomorphic 3-form on $X_6$;
\end{enumerate}
\item $F$ satisfies the Hermitian Yang-Mills equations:
$
\omega^{\mu \overline{\nu}} F_{a\overline{\nu}} = 0 \ ,
F_{\mu \overline{\nu}} = F_{\overline{\mu}\overline{\nu}} = 0 \ .
$
\end{enumerate}
In the above,  $\wedge$ is the wedge-product of differential forms and $\partial$ is the derivative with respect to the holomorphic coordinate (appropriately generalized for forms).
The general solutions to this system continue to inspire research today, engendering more geometric structures that contribute to the landscape.

Technically, the field strength lives on a vector bundle $\cV$ on $X$ and the case discussed at the beginning of this section, of $X$ being CY$_3$ and $\cV = T_X$ came to be known as the ``standard embedding'' and gave rise to $E_6$ GUT theories.
Non-standard embedding, by taking the vector bundle $\cV$ to have structure group $SU(4)$ or $SU(5)$ gives $SO(10)$ and $SU(5)$ GUTs, respectively. 
Equations \eqref{gens} and \eqref{TX3} generalize to so-called computations of {\it bundle cohomology groups} of $\cV$, $\cV^*$ and their tensor powers, which give the low-energy particle spectrum. 
The cubic Yukawa couplings in the Lagrangian constituted by these particles (fermion-fermion-Higgs) are appropriate tri-linear maps \footnote{This works out perfectly for a Calabi-Yau 3-fold: for example, $H^1(X,\cV) \times H^1(X,\cV) \times H^1(X,\cV) \rightarrow H^3(X,\cO_X) \simeq \IC$.} 
taking these cohomology groups to $\IC$.
The group theory \footnote{
The corresponding cohomology groups are as follows.
For $SU(5)$ we have
\[
\begin{array}{lll}
\#(\ol{10}) = h^1(X,\cV), & \#(10) = h^1(X,\cV^*), & \#(5) = h^1(X,\av), \\
\#(\ol{5}) = h^1(X, \av^*), & \#(1) = h^1(X,\vv), & \#(24) = 1,
\end{array}
\]
where the MSSM quarks/lepton live in the $10 \oplus \ol{5}$, and the
Higgs, in the $5$.
Similarly, for $SO(10)$, we have
\[
\begin{array}{lll}
\#(\ol{16}) = h^1(X,\cV^*), & \#(16) = h^1(X,\cV), & 
\#(10) = h^1(X,\av) = h^1(X, \av^*), \\
& \#(1) = h^1(X,\vv), & \#(45) = 1,
\end{array}
\]
where the SM quarks/lepton live in the 16, and the Higgs, in the 10.
} can be summarized as:
\[
\begin{array}{|l|ccl|}\hline
	E_8 \rightarrow G \times H & && \mbox{Breaking Pattern} \\ \hline
	SU(3) \times E_6 &
	248 & \rightarrow& (1,78) \oplus (3,27) \oplus
	(\overline{3},\overline{27}) \oplus (8,1) \\
	SU(4) \times SO(10) &
	248 & \rightarrow& (1,45) \oplus (4,16) \oplus
	(\overline{4},\overline{16}) \oplus (6,10) \oplus  (15,1) \\
	SU(5) \times SU(5) &
	248 & \rightarrow& (1,24) \oplus (5,\overline{10}) \oplus
	(\overline{5},10) \oplus
	(10, 5) \oplus (\overline{10},\bar{5}) \oplus  (24,1) \\
	\hline
\end{array}
\]

Constructing bundles over CY$_3$ satisfying the Hull-Strominger system has become an industry of realistic model building (q.v.~a long programme initiated by Ovrut et al.~\cite{guts}).
As mentioned at the end of the last subsection, one could proceed one more step by Wilson line projection from the GUT to the SM \footnote{
Geometrically, this means we need to find a Calabi-Yau 3-fold $X'$ with freely-acting discrete group $\Gamma$ so that the smooth quotient $X = X'/\Gamma$ has fundamental group $\Gamma=\pi_1(X)$ and admits a stable $\Gamma$-equivariant bundle $\cV$.}.
The first heterotic compactifications with {\it exact} MSSM particle content were constructed by finding appropriate bundles on the quotient of the Schoen manifold $S$, thereby answering the 20-year old challenge \cite{HetMSSM}.
Another was found over the bi-cubic from the CICY database in \cite{Anderson:2009mh}.
Exploring the ``zoo'' of small Hodge Calabi-Yau 3-folds which seems to be a fertile ground for exact SM solutions itself became a programme of Candelas et al.~\cite{Candelas:2007ac,smallhodge}.

All discrete, freely acting symmetries for all CICYs were classified using an impressive computer search using the computer algebra packages \cite{gap,sage} by Braun \cite{Braun:2010vc}.
Subsequently, a systematic scan over bundles (of the form of direct sum of line-bundles and with chosen K\"ahler parametres in the stability region) over all CICYs (up to $h^{1,1} = 6$) was performed by A.~Lukas et al., with the right cohomology groups computed, so that the result is the exact particle content of the SM \cite{CICYscan}.  This is a tremendous task and some 200 models were found in about $10^{10}$ candidates.
This is curiously in line with the 1-in-a-billion statistic from the type II scan of finding the SM \cite{Gmeiner:2005vz}.

Meanwhile, from a mathematical point of view, several generalizations of the CICY data was undertaken.
CICY4, the four-fold version of \eqref{cicy}  was completed in \cite{CICY4} and  921,497 were found.
Furthermore, it was found that one can in fact relax the condition that the configuration matrix entries be non-negative \cite{gCICY}, giving so-called gCICYs (generalized CICYs), whereby generating vast numbers of new CY$_3$s.

%%%%%%%%%%%%%%%%%
\subsection{A Plethora of CY$_3$}

Indeed, this brings us to the heart of a question, in essence both mathematical and physical: {\it how many CY$_3$s are there}?
In a way, we have an interesting sequence: in complex dimension 1, there is only the torus which is CY$_1$, in dimension 2, there turn out to  2 (the 4-dimensional torus $T^4 = (S^1)^4$ and a so-called K3 surface).
Starting in dimension 3, we till this day still have no idea how many distinct smooth manifolds are there, even though we have found literally billions as we shall now see.
In this section we will present the various databases constructed since the CICYs, which progressed in approximate 5-year periods since the late 1980s. 

%%%%%
\subsubsection{Weighted Hypersurfaces}
After the success story of CICYs, the search continued, partially due to the fact that all $\chi = 2 (h^{1,1} - h^{2,1})$ for CICYs were negative, but it was already suspected there should be {\bf mirror symmetry} so that each CY$_3$ with Hodge pair $(h^{1,1}, h^{2,1})$ has a mirror with these numbers reversed.
Thus there should be, for each negative $\chi$, another CY$_3$ with positive $\chi$.
Mirror symmetry  is one of the most exciting subjects of modern mathematics and theoretical physics, into which we unfortunately cannot delve due to limitations of space and we refer the readers to the authoritative volumes of \cite{mirror}.

Recall that our first example of the quintic $Q$ had $\IC\IP^4$ as its ambient space $A$.
Another natural generalization is to take {\it weighted} projective space $\IC\IP^4_{[d_0:\ldots:d_4]}$ as $A$; this generalizes \eqref{cpn} by having integer ``weights'' $(d_0,d_1,d_2,d_3, d_4) \in \IZ_{+}$ as
\begin{equation}\label{wp4}
\IC\IP^4_{[d_0:\ldots:d_4]} := \IC^5 \backslash \{ \vec{0} \} \bigg/ \left( (z_0,z_1,\ldots,z_4) \sim (\lambda^{d_0} z_0,\ldots,\lambda^{d_4} z_4) \right)\ , \qquad
\lambda \in \IC \backslash \{0\}\ . 
\end{equation}
Of course, taking all weights $d_i=1$ is the ordinary $\IC\IP^4$.
As with $Q$, if we embed a hypersurface of degree $d_0 + d_1 + \ldots + d_4$ into $\IC\IP^4_{[d_0:\ldots:d_4]}$, it defines a CY$_3$.

One caveat is that unlike ordinary projective space, weighted projective spaces are generically singular, and care must be taken to make sure the hypersurface avoids these singularities.
The classification of such manifolds was performed in \cite{Candelas:1989hd} and a total of 7555 is found, of which 28 have\footnote{Of course, with the importance of Wilson Lines, we should no longer be limited $|\chi| = 6$, but rather those with freely acting discrete groups of order $k$ and $\chi = \pm3k$.}
 Euler number $\pm6$.
In all, there are 2780 distinct Hodge pairs and with a more balanced $\chi \in [-960, 960]$.

%%%%%
\subsubsection{Elliptic Fibrations}
The mid 1990s saw the ``Second String Revolution'', and with the advent of dualities and branes which linked the various string theories, the traditional heterotic compactification scenario subsequently experienced a period of relative cool compared to its birth a decade earlier.
Due to the web of dualities, in particular that between the heterotic string and F-theory, there emerged another family of CY$_3$ studied in the 1990s;
this is the class of {\it elliptically fibred CY$_3$} \cite{elliptic}.

This is a generalization of our familiar CY$_1$, the torus, by allowing the elliptic curve to vary over a complex base surface.
Algebro-geometrically, this amounts to allowing the coefficients $g_2,g_4$ in the Weierstra\ss\ equation in \eqref{weierstrass} to depend on some appropriate coordinates of the base surface (or more strictly, to take values in dual of the canonical bundle of the base), arriving at an overall CY$_3$ as the total space.
What are the possible bases? Once again, this turned out to be a finite set \footnote{
Namely,
(I) Hirzebruch surfaces $\IF_r$ for $r = 0, 1, \ldots, 12$;
(II) So-called $\IC\IP^1$-blowups of Hirzebruch surfaces $\widehat{\IF}_r$ for $r=0,1,2,3$;
(III) Del Pezzo surfaces $d\IP_r$ for $r = 0,1,\ldots,9$;
and 
(IV) Enriques surface $\IE$.
These are classical complex surfaces which are the likes of $\IC\IP^1$ fibred over $\IC \IP^1$ as well as blowups of $\IC\IP^2$.
}
which was much explored in the 1990s \cite{Candelas:1997eh}, uncovering a wealth of elegant structure.

In light of F-theory model building, elliptic CY$_3$ and CY$_4$ have recently been investigated with renewed zest \cite{ellipticRecent}, and a new programme in identifying elliptically fibered CYs amongst the established databases has been ongoing with the help of modern computing \cite{ellipticClassify}.
It is widely believed that, in fact, {\em the large majority of CYn are elliptic fibrations over some appropriate $(n-1)$-fold}.
For example it was found in \cite{ellipticCICY}, among the CICYs, more than 99\% of the 7890 are elliptic fibrations; likewise, among the some $10^6$ CICY 4-folds, more than $99.9\%$ are elliptic.

%%%%%
\subsubsection{Toric Hypersurfaces: the KS Database}
So far, we have seen the ambient space $A$ of CY$_3$ being $\IC\IP^4$ (the quintic), as well as such generalizations to products of $\IC\IP^n$ (CICY) and weighted $\IC\IP^4$.  
One systematic generalization of weighted projective space is a {\bf toric variety}, which, instead of having a single list of weights as in \eqref{wp4}, has a list of $m$ weights (giving a so-called charge-matrix) acting on $\IC^{n+m}$ to give an $n$-fold.
Founded on the theoretic development of Batyrev-Borisov \cite{bb}, Kreuzer and Skarke spent almost a decade\footnote{Due to the untimely death of Max Kreuzer, it became a pertinent issue to attempt to salvage the data for posterity, a recent version of this legacy project is presented in \cite{newKS}.
One is also referred to \cite{flag} where the ambient space is more general than a toric variety} explicitly constructing such Calabi-Yau manifolds,
culminating in the early 2000s with the construction of the most extensive database of CY$_3$ so far, the {\bf Toric Hypersurfaces} \cite{KS}.

In brief, the ambient space is a toric 4-fold $A$, and the charge-matrix of weights can be repackaged into the vertices of an {\it integer polytope} (i.e., a convex body whose vertices are lattice points) living in $\IR^4$. This is the key to toric geometry: {\em to translate algebraic geometry to the combinatorics of integer lattices and polytopes}.

In particular, we focus on an integer polytope $\Delta \subset \IR^4$ which is {\bf reflexive}, meaning that $\Delta$ has a single interior point (which can be taken to be the origin) and all bounding hyperplanes are distance 1 from this point.
Equivalently, from $\Delta$ one can define \footnote{
In the perhaps more familiar definition of a toric variety in terms of fans of cones, the fan $\Sigma$ is simply the faces of $\Delta^{\circ}$.
} the {\it dual polytope} $\Delta^{\circ} := \left\{
\vec{v} \in \IR^4 | \vec{m} \cdot \vec{v} \ge -1 \ \forall \vec{m} \in \Delta
\right\}$.
Then, $\Delta$ is {\it reflexive} if $\Delta^\circ$ has also integer vertices. 
In such a case, a hypersurface in the toric variety $A$, constructed from the polytope data, is a CY hypersurface.

Specifically, the defining equation of the CY$_3$ is given by
\begin{equation}\label{bbHypersurface}
X = \{
\sum\limits_{\vec{m} \in \Delta} c_{\vec{m}} \prod\limits_{j=1}^k x_j^{\vec{m} \cdot \vec{v}_j + 1} = 0
\} \subset A \ ,
\end{equation}
with $x_j$ coordinates of the ambient toric 4-fold, $c_{\vec{m}}$ complex coefficients, and $\vec{v}_j$ the (integer) vertices of $\Delta^\circ$.
Weighted $\IC\IP^4$ (and of course ordinary $\IC\IP^4$) are archetypal examples of reflexive toric 4-folds.
For our familiar example of $Q \subset \IC\IP^4$,  $\Delta = {\tiny \mat{
-1 & 4 & -1 & -1 & -1 \\
-1 & -1 & 4 & -1 & -1 \\
-1 & -1 & -1 & 4 & -1 \\
-1 & -1 & -1 & -1 & 4 \\
}}$ and $\Delta^\circ = {\tiny \mat{
1 & 0 & 0 & 0 & -1 \\
0 & 1 & 0 & 0 & -1 \\
0 & 0 & 1 & 0 & -1 \\
0 & 0 & 0 & 1 & -1 \\
}}$ where the columns are the vertices of the polytopes, and one can check that the dot product of each column of one with that of the other is $ \geq -1$, in accord with the definition.
Equation \eqref{bbHypersurface} is precisely the quintic hypersurface and the ambient toric variety here is exactly $\IC\IP^4$.

\begin{figure}[h!!!]
\begin{center}
  $\begin{array}{c}\includegraphics[trim=1mm 1mm 1mm 6mm, clip, width=5in]{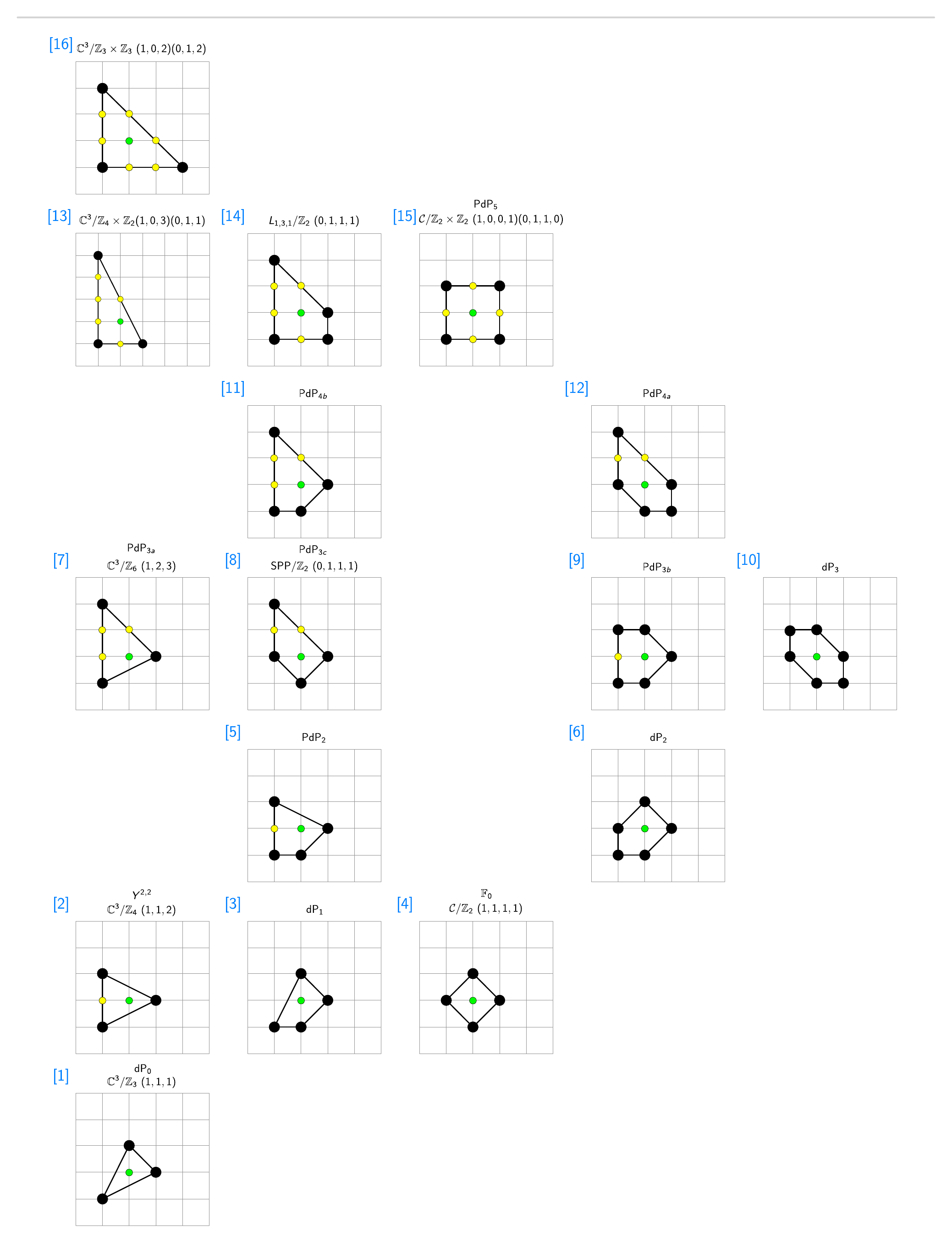}\end{array}$
\end{center}
\caption{
{\sf 
The 16 reflexive polygons $\Delta_2$. The single interior point (origin) is marked green, the vertices, black, and lattice points on the edges but are not vertices, in yellow (see \cite{He:2017gam}).
}
}
\label{f:16delta2}
\end{figure}

Thus the question of finding toric hypersurface CY$_3$ is the classification of reflexive integer 4-polytopes (up to redefinition by $SL(4; \IZ)$, under which the toric 4-folds are equivalent).
In $\IR^{1}$, there is trivially 1 reflexive polytope (the pair of points $\pm 1$).
In $\IR^2$, it is classically known that there are 16 reflexive polygons up to integer linear transformations with unit determinant (which give equivalent toric varieties).
For reference, we draw these in Figure \ref{f:16delta2};
hypersurfaces in the toric 2-folds created from these gives us 16 distinguished CY$_1$ as elliptic curves/tori.

Reflexive polytopes in $\IR^{n \geq 3}$ were already unknown to mathematicians until the work of Kreuzer-Skarke.
They found 4319 in $\IR^3$ and the computational challenge was to find all reflexive integer polytopes in $\IR^4$.
The actual calculation was performed on an SGI origin 2000 machine with about 30 processors (quite the state of the art in the 1990s) which took approximately 6 months; 473,800,776 was found.
Each of these gives a hypersurface CY$_3$ and thus from the database of tens of thousands established by the early 1990s, the list of CY$_3$ suddenly grew, with this tour de force, to half a billion. 

It should be emphasized that most of the ambient spaces $A$ (as with weighted $\IC\IP^4$) from the 500 million polytopes are {\it not smooth}, and requires smoothing or resolution of singularities: different resolutions give rise to potentially different CY$_3$s. 
Interestingly, in this large family, only 125 have smooth ambient space $A$ and more remarkably, only 16 have non-trivial fundamental group. Though of course a classification of discrete freely-acting symmetries has yet to be systematized, from which one could potentially extract many more non-simply-connected CY$_3$ by quotienting, these special 16 are quite interesting \cite{He:2013ofa}.

Now, while for a given $\Delta$, the Hodge pair will be the same, different resolutions will give different intersection numbers and Chern classes.
Up to $h^{1,1} = 7$, this was done exhaustively in \cite{Altman:2014bfa}, while for the highest $h^{1,1} \sim 490$, this was done in \cite{Braun:2017nhi}. The full list of CY$_3$s, after all the resolutions, has been recently estimated to be as large as $10^{10^5}$ \cite{Altman:2018zlc}.
In any case, the KS dataset  produced 30,108 distinct Hodge pairs and $\chi \in [-960, 960]$.
The extremal values of $\pm 960$ are actually the weighted $\IC\IP^4$ cases.
No CY construction so far has ever produced an Euler number whose magnitude exceeds 960.

\begin{figure}[h!!!]
\begin{center}
(a)
\includegraphics[scale=0.2]{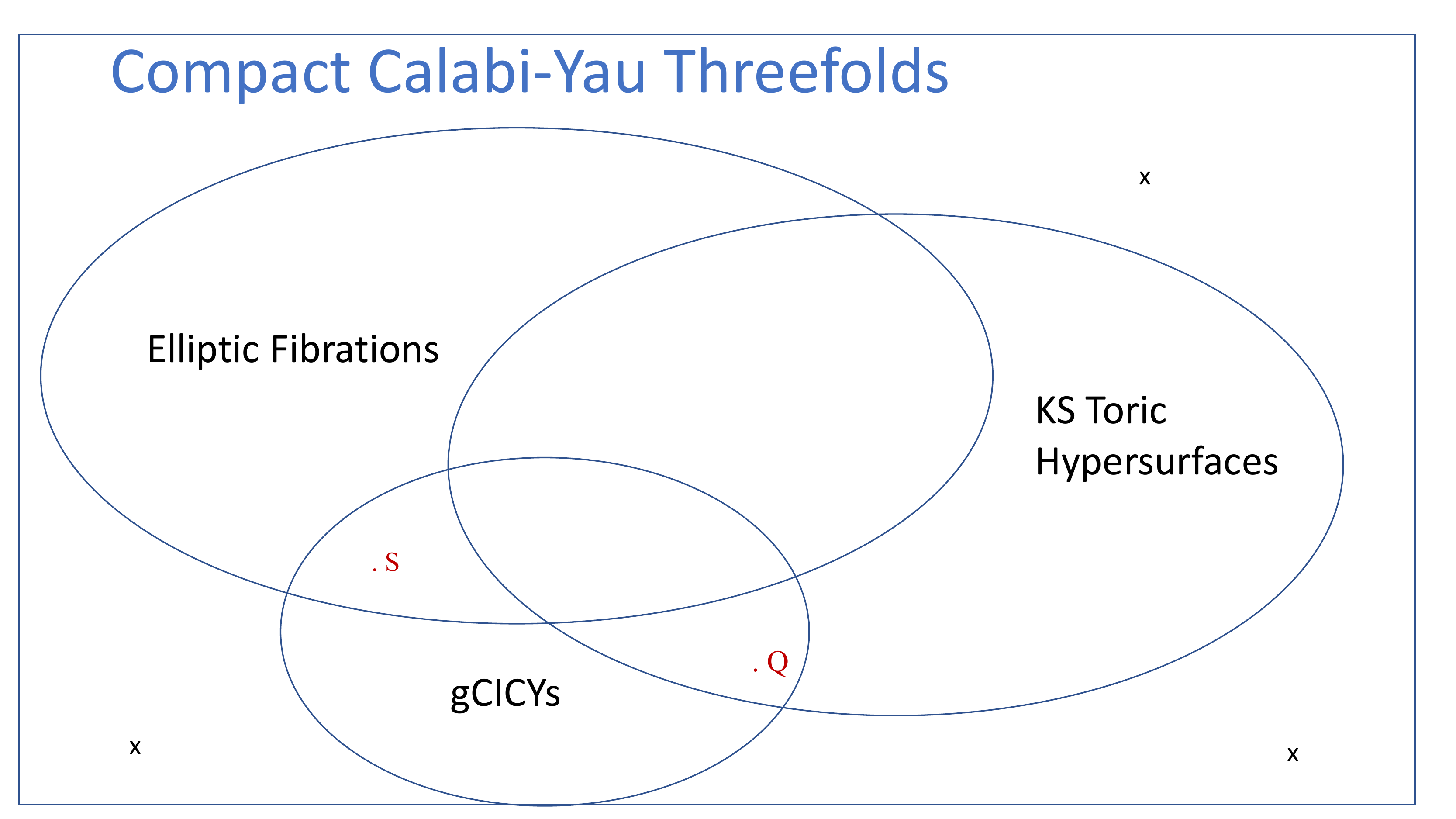}
(b)
\includegraphics[scale=0.5]{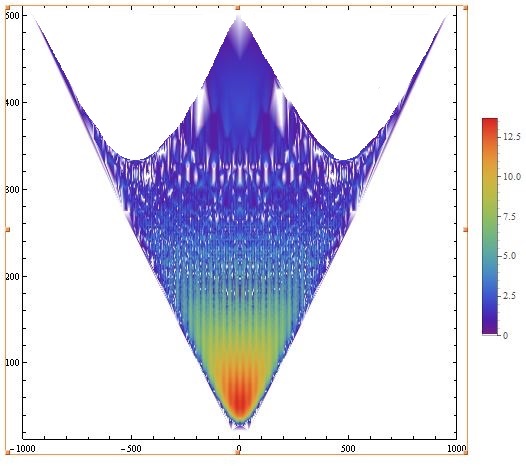}
\caption{{\sf 
(a) The space of CY$_3$, with the 3 most studied datasets.
There are also some individualized constructions outside the three major databases, symbolically marked as crosses.
$Q$ is the quintic, $S$ is the Schoen CY$_3$ and the most ``typical'' CY$_3$ has Hodge numbers $(27,27)$, totaling almost 1 million.
(b) Accumulating $\chi = 2(h^{1,1}-h^{1,2})$ (horizontal) versus $h^{1,1} + h^{1,2}$ (vertical) of the KS dataset in a colour Log-density plot (in a black-and-white rendition, the log-density can be seen from the intensity of the gray-scale).}}
\label{f:CY$_3$}
\end{center}
\end{figure}
%
%%%%%%%%%%%%%%%%%%%%%%%%%%
\paragraph{A Statistical Plot: }
There is an iconic plot: we draw the the KS dataset with $h^{1,1}(X) + h^{2,1}(X)$ in the vertical versus $\chi = 2(h^{1,1}(X) - h^{2,1}(X))$ in the horizontal
This is shown in part (b) of Figure \ref{f:CY$_3$}.
In part (a), we indicate our datasets discussed so far in a Venn diagram.
Several properties are of note. 
Since there are only a total of 30,108 distinct points, the some half-billion CY$_3$ are severely degenerate in $(h^{1,1},h^{1,2})$.
The funnel shape delineating the lower extremes is just due to our plotting difference versus sum of the (non-negative) Hodge numbers.
The fact that the figure is left-right symmetric is perhaps the best ``experimental'' evidence for mirror symmetry: that to each point with $\chi$ there should be one with $-\chi$, coming from the inter-change of the two Hodge numbers.
There is a paucity of CY$_3$ near the corners: near the very bottom tip and the top of the funnel shape, while a huge concentration resides near the bottom center (note this is a log-density plot).
In fact, the most ``typical'' CY$_3$ thus far known is one with Hodge numbers $(27,27)$, numbering about 1 million.
The distribution of the Hodge numbers is studied in detail in \cite{He:2015fif}.

%%%%%
\subsection{Finiteness?}\label{s:finite}
We have seen a multitude of CY$_3$s, with astronomical numbers.
Now, as seen from Figure \ref{f:RiemannSurface}, the topological type of Riemann surfaces (K\"ahler manifold of complex dimension 1) is determine by a single integer, the Euler number, the analogue in complex dimension 3 is given by a theorem of Wall \cite{wall} that the topology of a K\"ahler 3-fold is determined by
(1) the Hodge numbers; 
(2) the triple intersection numbers $d_{rst}$; and
(3) the first Pontrjagin class $p_1(T_M) = c_1(T_M)^2 - 2 c_2(T_M)$, where $c_2$ is the second Chern class to which we alluded in \S\ref{s:classic}.
For Calabi-Yau 3-folds, this amounts to the list of integers
\begin{equation}\label{CYdata}
\big\{ (h^{1,1}, h^{2,1}) \ ; \quad \ [c_2]_r \ ; \quad d_{rst} \big\}  \ , \qquad r,s,t = 1, \ldots, h^{1,1}  \ .
\end{equation}
This list should completely characterizes the topology.

In the early days, it was conjectured by Yau that the number of possible values for \eqref{CYdata} is finite for CY$_3$ (or indeed for Calabi-Yau manifolds of any dimension).
It is furthermore a conjecture of Miles Reid - the Reid Fantasy - that all Calabi-Yau manifolds are connected via topology-changing processes.

%%%%
\subsection{Flux Compactifications and Beyond}
It is worth re-iterating the point that CY$_3$s and even CY$_3$s with vector bundle data, as abundant as they are, comprise a small corner of the string landscape.
Indeed, even as far back as the mid-1980s, the number of vacua was already expected to be astronomical from the context of lattices \cite{Lerche:1986ae} (indeed, one of the authors, Schellekens, speculated upon the ``landscape'' some time before its popularity today \cite{Schellekens:2006xz}).

The often-quoted $10^{500}$ vacua first came from a rough estimate by M.~Douglas \cite{Douglas:2004zg} from the flux compactification scenario of \cite{flux}. In brief, one turns on various form-fields (generalizing the 1-form of electromagnetism) and integrates them on cycles (counted by our familiar Betti numbers) within Calabi-Yau spaces, giving us effective Lagrangians in terms of these ``fluxes'' (see \cite{Balasubramanian:2005zx}).
Thus, one can arrive at an approximate number of vacua, for example, from the typical size of the Betti numbers in known CY$_3$s.

We mention in passing that everything we have discussed are compact CYs.
Much like Liouville's theorem that bounded entire functions are constant, once we relax the condition of compactness, finiteness in classification is no longer expected.
Indeed, the landscape of non-compact CY$_3$s is infinite and
in the context of the AdS/CFT correspondence \cite{Maldacena:1997re}, such spaces provide the transverse direction for D-brane world-volume gauge theories which are of potential phenomenological interest. \footnote{
Geometrically, these non-compact CY$_3$ spaces can be seen as cones over Sasaki-Einstein manifolds \cite{SE} of real dimension 5.}

The simplest example is the trivially Ricci-flat $\IC^3$ (as an affine variety).
Any (singular) quotient of $\IC^3$, by discrete finite subgroups of $SU(3)$, is also locally CY$_3$.
More importantly, any lattice cone (in that the generators are lattice vectors) in $\IR^3$ with co-planar generating vectors gives a toric variety which is a non-compact CY$_3$ space. In other words, any lattice polygon gives an affine toric CY$_3$. Thus, we readily have an infinite family.
For instance, the 16 reflexive polygons in Figure \ref{f:16delta2} are perfect examples of non-compact Calabi-Yau 3-folds
(e.g., number [1] corresponds to a $\IZ_3$ quotient of $\IC^3$).
The mapping between the world-volume gauge theory and the polytope data of is a beautiful one, involving the combinatorics of dimers, the physics of brane-tilings and the algebraic geometry of singularity resolutions \cite{dimers}.

In addition to geometrical compactifications, be they the CY$_3$ scenarios discussed in this article or (in theory duality-equivalent) scenarios such as F-theory compactifications on 4-folds,  M-theory compactification on $G_2$ manifolds etc., there are still non-geometrical situations where one only considers the world-sheet conformal field theory such as Gepner models. 
All these have brought the string landscape to such a size \footnote{For instance,
a recent estimate of consistent F-theory flux compactifications numbers as high as $10^{10^5}$ by Taylor-Wang \cite{Taylor:2015xtz}.
Even models with exact SM particle content are estimated to be very high powers of ten \cite{Constantin:2018xkj,Cvetic:2019gnh}.
} that a re-examination of whether all effective field theories coupled to gravity can have such UV completions as string theory is investigated in the so-called {\it Swampland conjectures} by C.~Vafa et al.~\cite{swamp}.

%%%%%%%%%
\section{Conclusions \& Prospectus}
With the ever-growing number of Calabi-Yau structures, the number of ways to reach phenomenologically viable 4-dimensional effective theories is staggering.
In some sense, string theory has traded one difficult problem, quantization of gravity, with another, {\bf vacuum selection}.
The approach over the last few decades has been, as discussed above, to (1) establish concrete databases; (2) take advantage of the latest development in computer algebra \cite{m2,gap,singular,bertini,sage,mathematica,siam} and exhaustively calculate the relevant physical quantities (such as bundle cohomology for particle content); and (3) sift for vacua akin to our universe.

In the end, when all the constraints such as the right Yukawa coupling and masses, etc., have been enforced, it could still be the case that our universe is very special and rare (q.v.~\cite{Candelas:2007ac}).
However, confronted with the enormity of the landscape, this exhaustive search is computationally unfeasible, especially given that the key algorithms, be they Gro\"obner bases, which are at the heart of computational algebraic geometry or lattice polytope triangulations, which are at the core of combinatorial geometry, are known to be {\it exponentially expensive}.
Thus, statistical and data-science methods \cite{Douglas:2003um,Conlon:2004ds,Cicoli:2013zha,Dienes:2006ut,Denef:2006ad,Cole:2018emh} appear to be the way forward.

%%%
\subsection{Deep Learning the Landscape}
In \cite{He:2017aed}, a paradigm was proposed to attempt to use machine-learning (we will loosely refer to the collection of techniques such as neural networks, support vector machines, or decision trees, etc., as simply ML) to bypass the aforementioned standard algorithms, in particular to study the string landscape and beyond.
It was found that central problems such as computing cohomology of vector bundles appear to be machine-learnable to very high precision. 
Indeed, \cite{He:2017aed,Ruehle:2017mzq,Krefl:2017yox,Carifio:2017bov} brought machine learning to the string landscape in 2017 and there has been a host of activity since.

The basic idea is that whatever the problem at hand, in algebraic or combinatorial geometry, we typically compute an quantity (usually a positive integer) $O$ from a given configuration (usually some integer matrix) $I$, using sophisticated and computationally expensive methods from modern mathematics.
This has been done by brute force over the decades, with the help of computer algebra to establish datasets of the form $\cD = \{I_i \to O_i\}_{i=1,2,\ldots,N}$ for a large number $N$ of cases.
Such a situation is perfectly adapted for {\bf supervised machine-learning} \footnote{
Likewise, one could have an incomplete classification of configurations $\{I_i\}$, and attempt to extrapolate to the remaining by unsupervised learning.
} (the reader is referred to the canonical introduction \cite{mitchell}).

At the fundamental level, supervised learning by ML is simply an intricate form of (non-linear) regression:  (1) a host of internal parameters and a directed graph of nodes, each of which is some function; (2) these are fitted to available data in order to specify the parameters by minimizing an appropriate cost-function (such as sum of errors squared), and (3) prediction and extrapolation are then attempted.
To avoid over-fitting and to see how well the ML is performing as there  could be literally thousands of parameters, the canonical method is {\it cross-validation}.
The data $\cD$ disjointly into a training set and a validation set: $\cD := \cT \sqcup \cV$. We optimize the parameters in the ML on $\cT$ and check how it performs on $\cV$, which it has not ``seen'' before.
Usually, one steps gradually in the percentage of $\cT$ in $\cD$ (with a few rounds of random samples to estimate error bars) and plot the accuracy measure, giving us a so-called {\bf learning curve}.
To entice the reader, let us illustrate with two case studies.

\begin{figure}[h!!!]
\begin{center}
(a)
\includegraphics[scale=0.4]{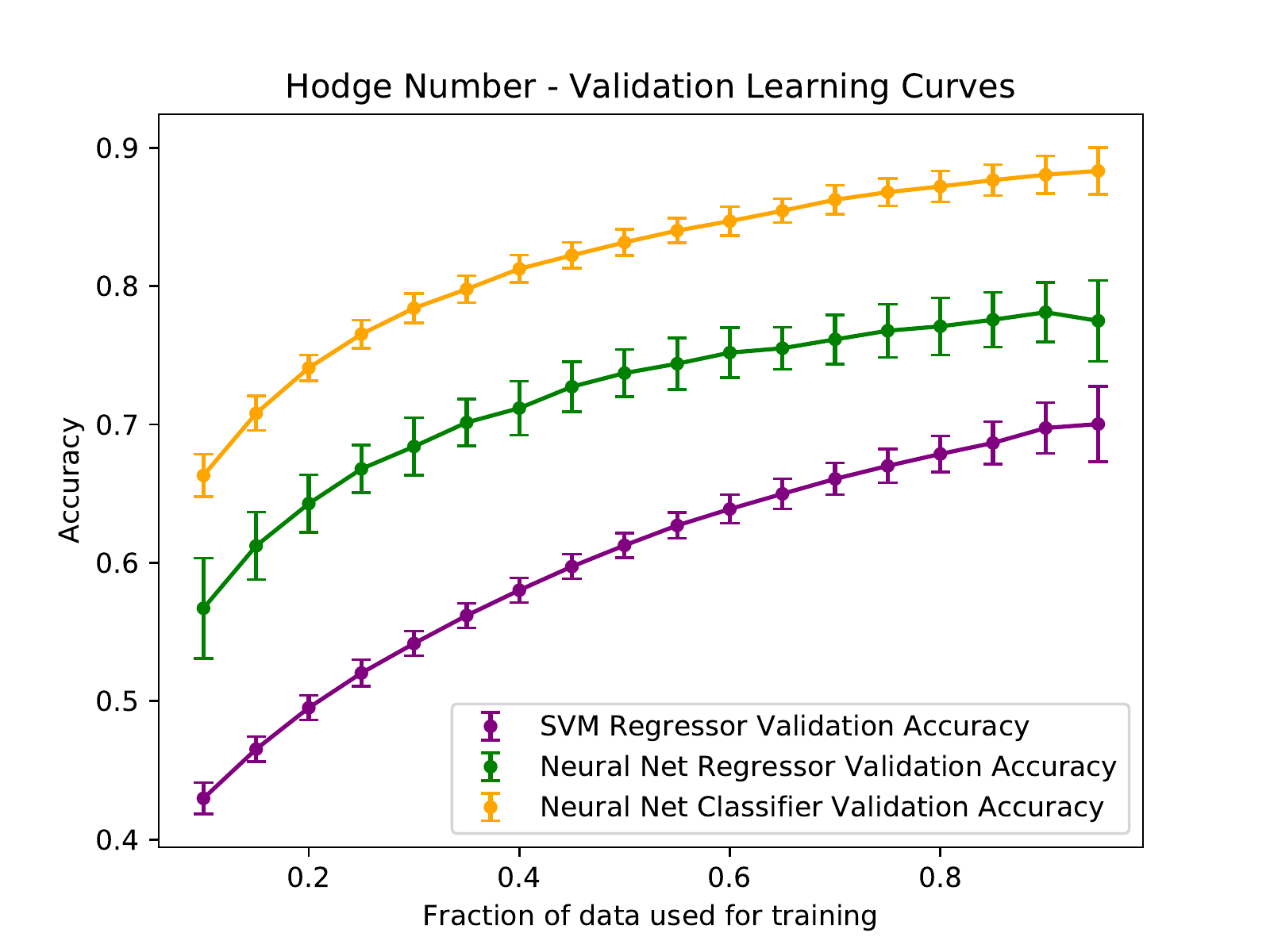}
(b)
\includegraphics[scale=0.6, trim=0mm 0mm 40mm 0mm, clip]{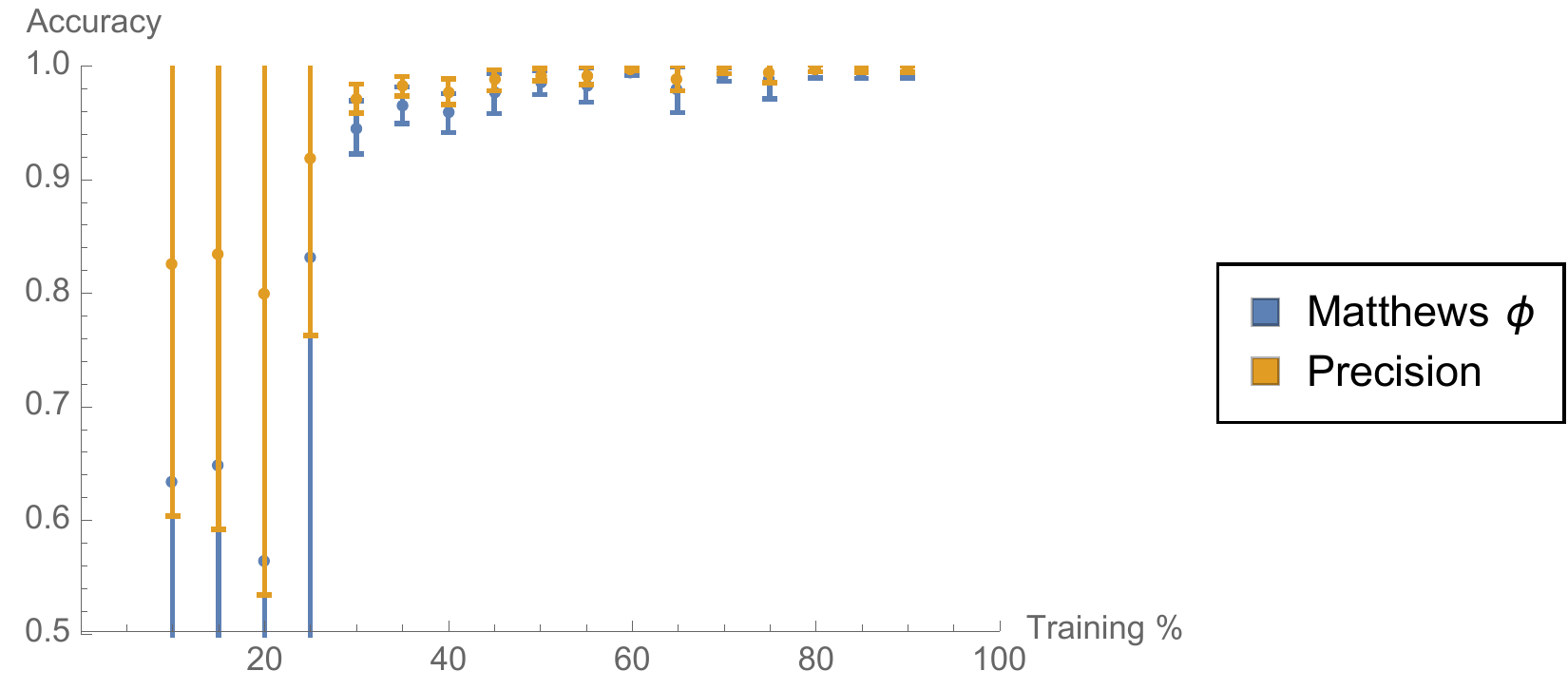}
\caption{{\sf 
(a) The learning curve for the precision of 3 machine-learning methods on learning the exact Hodge number $h^{1,1}$ of CICY 3-folds.
(b) The learning curve of precision (in yellow) and Matthew's $\phi$-coefficient (blue) for recognizing whether a CICY is elliptically fibered using a SVM.
All error bars come from collecting 10 random samples from the training set.
}}
\label{f:MLeg}
\end{center}
\end{figure}

\paragraph{Learning Hodge Numbers: } 
Consider the CICYs. Whilst the Euler number is easily computed from the configuration matrix analytically (see.~\cite{hubschbook}), the individual terms 
$(h^{1,1},h^{2,1})$ involve exact sequence chasing, in particular determining the (co-)kernels of the maps in the sequence which very quickly becomes of high dimension.
This was nevertheless performed in \cite{cicy} over the course of the years.
Can ML ``guess'' at the right value of $h^{1,1}$ by simply ``looking'' at the configuration matrix? In principle, there might be some form of an equation of $h^{1,1}$ in terms of $q_r^j$, but it would involve so many conditionals on how $q^r_j$ fits into an induced long exact sequence in cohomology that such equations would be of tremendous length and rather unhelpful.

In \cite{He:2017aed,cicyML}, this exercise of pattern-recognition of $h^{1,1}$ from $q_r^j$ was done and the learning curve is shown in Figure \ref{f:MLeg} (a).
We compare 3 different methods, a neural network, a support vector machine and a neural classifier.
The accuracy measure is the percentage in $\cV$ of when the $h^{1,1}$ agrees between the actual and the predicted (note that among the 7890 manifolds, $h^{1,1}$ can range from 1 to 19).  We see that the classifier performs the best and at 50\% training data, we are already getting over 80\% correct prediction for $h^{1,1}$. The error bars are due to different random sampling of $\cT$. Crucially, instead of taking the many hours to compute exactly, ML is guessing the result in a matter of seconds.

\paragraph{Learning Elliptic Fibrations: }
Continuing with the CICYs as a playground, a substantial fraction of which are, in fact, elliptically fibred.
By an exhaustive search using theorems and conjectures of Oguiso, Koll\'ar, Wilson et al., \cite{ellipticThm}, 
the elliptically fibred ones were distinguished \cite{ellipticCICY}.
Can ML do this identification as a 1/0 binary query (yes/no to elliptic fibration) without any knowledge of the underlying mathematics but merely by ``looking'' at the configuration matrix?
Again, there is no known simple formula of the 1/0 in terms of $q_r^j$, nor is there expected to be one.

Using a support vector machine which finds an optimal hyperplane in the very high-dimensional vector space whose entries are the flattened configuration matrix (considering the maximum number of rows and columns of $q_r^j$, this is $\IZ^{12 \times 15}$) between 0/1, the learning curve was constructed in \cite{He:2019vsj}.
This is shown in Figure \ref{f:MLeg} (b).
We plot both the precision in yellow (percentage of 1/0 agreed between actual and predicted) as well as so-called Matthew's $\phi$-coefficient in blue (this is  a scaled version of the chi-squared which is well adapted to binary classification problems in order to account for false-positives; the closer $\phi$ is to 1, the better the confidence of the agreement).
We see that even at as small as 30\% training data, ML can predict which are elliptically fibred to 95\% precision and confidence over the entire CICY dataset  \footnote{
One might worry that the elliptic dataset is highly imbalanced in that the vast majority are elliptically fibred; we ensured balanced data by enhancing the positives with permutations, which are equivalent representations of the same manifold.
In other words, the ML is not predicting 1 all the time.
}.
Again, the guessing is done in seconds rather than the many hours of computation.

\subsection*{Epilogue}
We have taken a promenade in the space of CY$_3$ in the context of the string landscape, in a more or less historical order of the appearance of the constructions.
From the initial triadophiliac query of whether there exists smooth compact CY$_3$ with Euler number $\pm6$, to the proliferation of CY$_3$s constructed from algebro-geometric and combinatorial methods, to the first exact SM particle content from a CY$_3$ endowed a vector bundle that solves the Hull-Strominger system, to the clear future of applying data-scientific and ML techniques, our journey has spanned some four decades.
Throughout, we have been mindful of the intricate interplay between the mathematics and physics, emboldened by the plenitude of data and results, and inspired by the glimpses towards the yet inexplicable.
The cartography of Calabi-Yau manifolds will certainly continue to provoke further exploration, in an ever-growing inter-disciplinary research in  physics, mathematics, and computer science/artificial intelligence.

\section*{Further Reading}
The interested reader is referred to the following reviews, lecture notes and books for further reading:
\begin{itemize}
\item
Candelas \cite{Candelas:1987is} for a rapid introduction of algebraic geometry for physicists;
the classic book on CY manifolds by H\"ubsch \cite{hubschbook};
a survey of CY by Yau \cite{yaubook2};
lecture notes on CY and string theory by Greene \cite{Greene:1996cy} and Aspinwall \cite{Aspinwall:2004jr};
a more recent compendium by Gross-Huybrechts-Joyce \cite{grossbook};

\item 
Ib\'{a}n\~{e}z-Uranga \cite{IU}, Gra\~{n}a \cite{Grana:2005jc} for string phenomenology and flux compactifications on CY as well as counting string vacua by Denef \cite{Denef:2008wq};  

\item
the classic textbooks on algebraic geometry by Griffith-Harris \cite{GH}, Hartshorne \cite{hart},
as well as an invitation to computational algebraic geometry by Schenck \cite{m2book};

\item
the standard introduction to machine-learning by Mitchell \cite{mitchell} as wel as  \cite{He:2018jtw} for a pedagogical introduction for machine-learning CY.
\end{itemize}

\let\oldthebibliography=\thebibliography
\let\endoldthebibliography=\endthebibliography
\renewenvironment{thebibliography}[1]{%
\begin{oldthebibliography}{#1}%
\setlength{\parskip}{0ex}%
\setlength{\itemsep}{0ex}%
}%
{%
\end{oldthebibliography}%
}

\end{document}